# The orthometric parameterisation of the Shapiro delay and an improved test of general relativity with binary pulsars

Paulo C. C. Freire[1]⋆ & Norbert Wex[1]
[1]*Max-Planck-Institut für Radioastronomie, Auf dem Hügel 69, D-53121 Bonn, Germany*

**ABSTRACT**
In this paper, we express the relativistic propagational delay of light in the space-time of a binary system (commonly known as the "Shapiro delay") as a sum of harmonics of the orbital period of the system. We do this first for near-circular orbits as a natural expansion of an existing orbital model for low-eccentricity binary systems. The amplitudes of the 3$^{\rm rd}$ and higher harmonics can be described by two new post-Keplerian (PK) parameters proportional to the amplitudes of the third and fourth harmonics ($h_3, h_4$). For high orbital inclinations we use a PK parameter proportional to the ratio of amplitudes of successive harmonics ($\varsigma$) instead of $h_4$. The new PK parameters are much less correlated with each other than $r$ and $s$ and provide a superior description of the constraints introduced by the Shapiro delay on the orbital inclination and the masses of the components of the binary. A least-squares fit that uses $h_3$ always converges, unlike in the case of the $(r, s)$ parameterisation; its resulting statistical significance is the best indicator of whether the Shapiro delay has been detected. Until now these constraints could only be derived from Bayesian $\chi^2$ maps of the $(\cos i, m_c)$ space. We show that for low orbital inclinations even these maps overestimate the masses of the components and that this can be corrected by mapping the orthogonal $(h_3, h_4)$ space instead. Finally, we extend the $h_3, \varsigma$ parameterisation to eccentric binaries with high orbital inclinations. For some such binaries we can measure extra PK parameters and test general relativity using the Shapiro delay parameters. In this case we can use the measurement of $h_3$ as a test of general relativity. We show that this new test is not only more stringent than the $r$ test, but it is even more stringent than the previous $s$ test. Until now this new parametric test could only be derived statistically from an analysis of a probabilistic $\chi^2$ map.

## 1 INTRODUCTION

In 1964 Shapiro pointed out that an electromagnetic wave passing near a massive body, such as the Sun, suffers a relativistic time delay (Shapiro 1964). This was the fourth test of general relativity (GR), after the three classical tests proposed by Einstein: the perihelion precession of Mercury, the deflection of light by the Sun and the gravitational redshift (Einstein 1916). The experimental value for the Shapiro delay determined by the Cassini spacecraft agrees with the GR prediction at the 0.002% level (Bertotti et al. 2003).

Outside the Solar System, the Shapiro delay has been observed in a number of binary pulsars. If the inclination of the orbit with respect to the line-of-sight is sufficiently high then near superior conjunction the radio pulse will experience a measurable delay on its way from the pulsar to Earth (Lorimer & Kramer 2005). In this case, assuming GR is correct, the Shapiro delay allows to constrain or even determine the masses of the system. An early example of this is PSR B1855+09 (Ryba & Taylor 1991; Kaspi et al. 1994). In binary pulsars where additional relativistic effects are observed the Shapiro delay can be used to test GR (Stairs 2003); examples of this are PSR B1534+12 (Stairs et al. 2002) and the "double pulsar" PSR J0737−3039 (Kramer et al. 2006b). Although these tests have not reached the same level of precision as the Cassini experiment in the Solar System, they are important since they complement the Solar System experiments. In particular in systems where the companion is a neutron star, one could expect strong-field effects on the propagation of photons, which would not be measurable in the weak field of the Solar System (Damour & Taylor 1992; Damour & Esposito-Farèse 1996).

⋆ E-mail: pfreire@mpifr-bonn.mpg.de (PCCF)

In this paper we develop a new parameterisation of the Shapiro delay in binary pulsars based on the Fourier expansion of this effect in harmonics of the orbital period (§ 3). As we will show, this provides a superior description of the Shapiro delay (§ 4), provides a new space for improved $\chi^2$ maps (§ 5) and a much improved parametric test of GR (§ 6).

## 2  PULSAR TIMING AND SHAPIRO DELAY

Pulsar timing is the technique that makes some neutron stars superb astrophysical tools. It is especially useful in cases where the pulsar is located in a binary system (Stairs 2003 and references therein). In such cases, the orbital motion introduces a delay to the times of arrival of the pulses at the Solar System Barycenter given by (Damour & Taylor 1992):

$$\Delta = \Delta_{\rm R} + \Delta_{\rm E} + \Delta_{\rm S} + \Delta_{\rm A}, \quad (1)$$

where to first order $\Delta_{\rm R}$ is the "Newtonian" delay due to the light travel time across the Keplerian part of the orbit (henceforth the "Rømer" delay), $\Delta_{\rm E}$ is the Einstein delay, $\Delta_{\rm S}$ is the Shapiro delay, and $\Delta_{\rm A}$ is the aberration delay, which generally cannot be separated from the Rømer delay.

The two main orbital models being used for binary pulsar timing (Damour & Deruelle 1986; Lange et al. 2001) parameterise the Shapiro delay as a function of two post-Keplerian (PK) parameters, the "range" ($r$) and "shape" ($s$) parameters:

$$\Delta_{\rm S}(\varphi) = 2r \ln \left[ \frac{1 + e \cos \varphi}{1 - s \sin(\omega + \varphi)} \right] \quad (2)$$

where $\varphi$ is the true anomaly $e$ is the orbital eccentricity and $\omega$ is the longitude of periastron relative to ascending node.

For most theories of gravity (Damour & Taylor 1992; Will 1993) we have

$$s = \sin i. \quad (3)$$

where $i$ is the orbital inclination. If general relativity (GR) is correct then

$$r = T_\odot m_c. \quad (4)$$

where $T_\odot \equiv GM_\odot c^{-3} = 4.925490947 \mu$s is the the mass of the Sun in units of time and $m_c$ is the mass of the companion star.

The mass function of the binary ($f$) is given by:

$$f = \frac{(m_c \sin i)^3}{(m_p + m_c)^2} = \frac{n^2 x_p^3}{T_\odot}, \quad (5)$$

where $n = 2\pi/P_b$ is the mean angular velocity of the binary, $P_b$ is the observed orbital period, $x_p$ is observed the projected semi-major axis of the pulsar's orbit (normally indicated in light seconds), and $m_p$ is the mass of the pulsar.

Apart from $r$ and $s$ there are other PK parameters that can be determined from the timing of relativistic binary pulsars within the theory-independent parameterised PK approach (Damour & Deruelle 1986; Damour & Taylor 1992): the most important among these are the rate of advance of periastron ($\dot\omega$), the amplitude of the "Einstein" delay ($\gamma$) and the orbital period decay ($\dot P_b$). Assuming a particular theory of gravity, these parameters can be expressed as functions of $m_p$ and $m_c$ (Damour & Taylor 1992); eq. (4) is the simplest of these when GR is used to do the mass calculations.

If we measure two precise PK parameters and assume a particular theory of gravity, then their equations completely determine $m_p$ and $m_c$. If we measure more than two PK parameters the system becomes over-determined; in that case we can test the consistency of the gravitational theory used to calculate the masses.

## 3  FOURIER SERIES OF PROPAGATIONAL DELAYS FOR LOW ECCENTRICITIES

In what follows we concentrate only on the propagational delay, $\Delta_{\rm RS} \equiv \Delta_{\rm R} + \Delta_{\rm S}$ for low-eccentricity orbits. First, we discuss the Rømer delay for low eccentricities in the formalism of Lange et al. (2001). Second, we expand the Shapiro delay for small-eccentricity orbits as a function of orbital harmonics, i.e., sine and cosine waves with frequencies that are integer multiples of the orbital frequency.

### 3.1  The Rømer delay for low eccentricities

In this section, we essentially follow Lange et al. (2001). If we neglect terms of order $e^2$ the Rømer delay can be written as

$$\Delta_{\rm R} \simeq x_p \left( \sin \Phi + \frac{\kappa}{2} \sin 2\Phi - \frac{\eta}{2} \cos 2\Phi \right), \quad (6)$$

where terms which are constant in time are omitted and

$$\Phi = n(T - T_{\rm asc}), \quad (7)$$

is the celestial longitude of the binary, $T_{\rm asc}$ is the *time of ascending node* and

$$\eta \equiv e \sin \omega \quad \text{and} \quad \kappa \equiv e \cos \omega, \quad (8)$$

are the *first* and *second Laplace-Lagrange parameters*. For the moment we consider these quantities to be constant in time. Their variation, and how it translates as a variation of the Keplerian parameters $x_p, e$ and $\omega$ is discussed in Lange et al. (2001).

### 3.2  Fourier expansion of the Shapiro delay for nearly circular orbits

For small-eccentricity binary pulsars, the Shapiro delay (eq. 2) can be re-written as follows:

$$\Delta_{\rm S} = -2r \ln(1 - s \sin \Phi) \equiv -2r f(\Phi). \quad (9)$$

The function $f(\Phi)$ can be expanded in a Fourier series:

$$f(\Phi) = \frac{a_0}{2} + \sum_{k=1}^{\infty} a_k \cos(k\Phi) + \sum_{k=1}^{\infty} b_k \sin(k\Phi), \quad (10)$$

with the following coefficients:

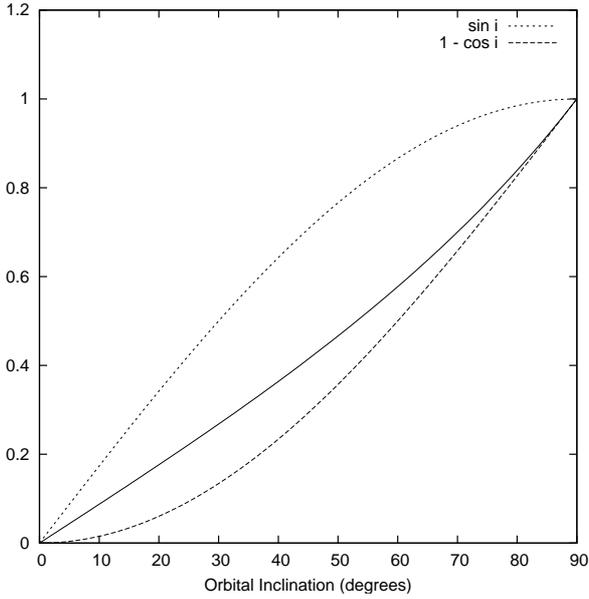

**Figure 1.** Behavior of $\varsigma$ (solid curve) with orbital inclination. For high orbital inclinations its variation with $i$ is similar to $1 - \cos i$.

$$
\begin{aligned}
a_0 &= 2 \ln\left(\tfrac{1+\bar{c}}{2}\right), \\
a_1 &= 0, & b_1 &= -2\left(\tfrac{s}{1+\bar{c}}\right), \\
a_2 &= \left(\tfrac{s}{1+\bar{c}}\right)^2, & b_2 &= 0, \\
a_3 &= 0, & b_3 &= \tfrac{2}{3}\left(\tfrac{s}{1+\bar{c}}\right)^3, \\
a_4 &= -\tfrac{1}{2}\left(\tfrac{s}{1+\bar{c}}\right)^4, & b_4 &= 0, \\
a_5 &= 0, & b_5 &= -\tfrac{2}{5}\left(\tfrac{s}{1+\bar{c}}\right)^5, \\
a_6 &= \tfrac{1}{3}\left(\tfrac{s}{1+\bar{c}}\right)^6, & b_6 &= 0, \\
\ldots & & \ldots &
\end{aligned} \quad (11)
$$

where $\bar{c} \equiv \sqrt{1-s^2} = |\cos i|$. We can define the "orthometric ratio" parameter to express the ratio of the amplitude of the successive harmonics of the Shapiro delay:

$$\varsigma \equiv \frac{s}{1+\bar{c}} = \left(\frac{1-\bar{c}}{1+\bar{c}}\right)^{1/2}. \quad (12)$$

Since $\varsigma$ depends solely on $s$, it also has a theory-independent meaning. Using $\varsigma$ one finds a closed-form expression for the Fourier coefficients in eq. (11):

$$
\begin{aligned}
a_0 &= -2\ln(1+\varsigma^2), \\
a_k &= (-1)^{\frac{k+2}{2}} \frac{2}{k} \varsigma^k, & k = 2,4,6,\ldots \\
b_k &= (-1)^{\frac{k+1}{2}} \frac{2}{k} \varsigma^k, & k = 1,3,5,\ldots
\end{aligned} \quad (13)
$$

These harmonics describe what we actually observe with pulsar timing (time delays) as a set of orthogonal functions (the orbital harmonics). Their orthogonality implies that their amplitudes are *a priori* uncorrelated. This has important implications, as we will discuss below.

### 3.3 Low-inclination case

If the orbital inclination is small, $s$ and $\varsigma$ will also be small and there will be a steep decrease in the power of each successive harmonic. We define as "low inclination" a case where harmonics higher than 2 are not detectable at the pulsar's timing precision, i.e., $\Delta_{RS}$ is given to sufficient precision by the sum of the first two harmonics $\Delta_{RS}^{(2)}$. Whether a binary pulsar's orbital inclination is "low" depends on its timing precision and the number of measured times of arrival (TOAs).

Ignoring constant terms, the Shapiro delay can then be described as:

$$\Delta_S \simeq -2r(b_1 \sin\Phi + a_2 \cos 2\Phi), \quad (14)$$

which corresponds to eq. (A17) in Lange et al. (2001). Adding eqs. (6) and (14) and (again) ignoring constant terms, we obtain for $\Delta_{RS}^{(2)}$ an equation similar to eq. (6):

$$\Delta_{RS}^{(2)} \simeq x_p^{obs}\left(\sin\Phi + \frac{\kappa}{2}\sin 2\Phi - \frac{\eta^{obs}}{2}\cos 2\Phi\right), \quad (15)$$

where

$$
\begin{aligned}
x_p^{obs} &= x_p - 2rb_1, & (16) \\
\eta^{obs} &= \eta + 4ra_2/x_p & (17)
\end{aligned}
$$

and we have used $x_p \ll r$. This implies that the first two harmonics of the Shapiro delay can be absorbed by a redefinition of $x_p$ and $\eta$. This means that for low inclinations the Shapiro delay cannot be separated from the Rømer delay. For a circular binary there will be an apparent eccentricity given by eq. (17) with $\omega = 90°$.

### 3.4 Medium-inclination case

If the sum of harmonics 3 and higher $\Delta_S^{(3+)}$ is not negligible then $\Delta_{RS}$ is given by

$$\Delta_{RS} = \Delta_{RS}^{(2)} + \Delta_S^{(3+)}. \quad (18)$$

For low orbital eccentricities harmonics 3 and higher are entirely due to the Shapiro delay. They can then be described by:

$$\Delta_S^{(3+)} = -4h_3\left(\frac{1}{3}\sin 3\Phi - \frac{1}{4}\varsigma\cos 4\Phi - \frac{1}{5}\varsigma^2\sin 5\Phi + \frac{1}{6}\varsigma^3\cos 6\Phi + \ldots\right), (19)$$

Because the first two harmonics of the Shapiro delay are absorbed by a re-definition of the Rømer delay (§ 3.3) $\Delta_S^{(3+)}$ represents the *measurable* part of the Shapiro delay; the new parameter

$$h_3 \equiv r\varsigma^3. \quad (20)$$

quantifies its amplitude, for that reason we call this the "orthometric amplitude parameter" of the Shapiro delay.

We define as "medium inclination" a binary pulsar where we can measure the amplitude of the 3rd harmonic ($4h_3/3$), but not the 4th and higher harmonics. The amplitude of the 4th harmonic is given by

$$h_4 \equiv h_3\varsigma; \quad (21)$$

its non-detectability implies that $\varsigma$ is significantly smaller than 1. That imposes an upper limit on $s$ (or a lower limit on $\bar{c}$), which are related to $\varsigma$ by:

$$s = \frac{2\varsigma}{1+\varsigma^2}, \quad \bar{c} = \frac{1-\varsigma^2}{1+\varsigma^2} \quad (22)$$

i.e., the binary cannot be edge-on, otherwise $\varsigma$ and $s$ would be close to 1 and harmonics higher than 3 should be measurable.

Therefore eq. (19) is truncated at $\sin 3\Phi$ and we can parameterise the Shapiro delay with $h_3$ only. Note that this quantity is always well determined; if the Shapiro delay is not detectable then $h_3$ will simply be consistent with 0 (within measurement errors). Furthermore, $h_3$ is not correlated with any of the parameters that describe the amplitude and phase of the lower harmonics ($x_p$, $T_{\rm asc}$, $\kappa$ and $\eta$), so its measurement (or lack of it) has no implications for any other parameters.

If $h_3 > 0$ we obtain a curve in the $r-s$ space (or $r-\bar{c}$ space) where the binary must be located from eq. (20):

$$r = h_3 \left( \frac{1+\sqrt{1-s^2}}{s} \right)^3 = h_3 \left( \frac{1+\bar{c}}{1-\bar{c}} \right)^{3/2}, \quad (23)$$

the uncertainty of $h_3$ turns this $r-s$ line into a band of finite width (we will henceforth refer to the region of any space that is consistent with the measurement of a given PK parameter and its 1-$\sigma$ uncertainty as its "1-$\sigma$ band"). This equation also means that for the same observed third harmonic amplitude (or $h_3$) there will be an infinity of equally valid $(r,s)$ combinations, although some solutions at high inclinations are excluded by the upper limit on $\varsigma$. This has the consequence that $r$ and $s$ will generally be highly correlated. Unless the orbital inclinations are high we cannot determine $s$, this means that $r$ cannot be determined with any useful precision.

### 3.5 High inclination case

We define an orbital inclination as "high" if harmonics above 3 are detectable. If harmonic 4 is detectable, then its amplitude is uncorrelated to $h_3$ and the Keplerian parameters. In principle this makes $(h_3, h_4)$ the best possible parameterisation of the Shapiro delay. As we will see below, this is not the case whenever many harmonics are detectable in the timing.

From each value of $h_4$ we obtain a curve in the $(s,r)$ space (or $(\bar{c}, r)$ space) given by

$$r = h_4 \left( \frac{1+\sqrt{1-s^2}}{s} \right)^4 = h_4 \left( \frac{1+\bar{c}}{1-\bar{c}} \right)^2. \quad (24)$$

From eqs. (21) and (22), we can see that this cuts the $h_3$ line at

$$s = \frac{2h_3 h_4}{h_3^2 + h_4^2}, \quad \bar{c} = \frac{h_3^2 - h_4^2}{h_3^2 + h_4^2} \quad (25)$$

$$r = \frac{h_3^4}{h_4^3}. \quad (26)$$

The high powers of $h_3$ and $h_4$ in the expression for $r$ mean that its uncertainty is always much larger than the uncertainties of $h_3$ and $h_4$. It is for this reason that measuring precise masses with the Shapiro delay is so difficult.

### 3.6 Very high inclination case

For $\sin i \simeq 1$ the ratio between successive harmonics (determined by $\varsigma$) is close to 1 and therefore we start detecting a large number of them. Up to order 4, the amplitude of each harmonic is uncorrelated and independent from the amplitudes of all previous harmonics. That is no longer the case for harmonics above 4 because the Shapiro delay can be parameterised by two parameters only (e.g., $h_3$ and $h_4$). Qualitatively, the amplitudes of the higher harmonics bring no new information.

So what is the benefit of measuring harmonics higher than 4? The time delays associated with them are given by:

$$\Delta_{\rm S}^{(5+)} = 4h_4 \left[ \frac{1}{5} \frac{h_4}{h_3} \sin 5\Phi - \frac{1}{6} \left( \frac{h_4}{h_3} \right)^2 \cos 6\Phi - \frac{1}{7} \left( \frac{h_4}{h_3} \right)^3 \sin 7\Phi + \ldots \right]. \quad (27)$$

This means that the amplitudes of the high harmonics contribute to a very precise measurement of the $h_4/h_3$ ratio ($\varsigma$). If this is measured with better precision than warranted by the uncertainty of $h_3$ and $h_4$ the latter parameters become correlated. If $|\rho(h_3, h_4)| > |\rho(h_3, \varsigma)| = 0.5$ (where $\rho(x,y)$ is the correlation between $x$ and $y$) then $h_3$ and $\varsigma$ provide the best description of the Shapiro delay. Thus for high inclinations we use $\varsigma$ instead of $h_4$ as the second independent parameter in the description of the Shapiro delay. As previously mentioned, $\varsigma$ has the advantage of measuring the orbital inclination in a theory-independent way, like the Shapiro "shape" parameter $s$. For high inclinations the variation of $\varsigma$ with $i$ is similar to that of $\cos i$ (see Fig. 1). For randomly oriented orbits $\cos i$ has constant probability density, therefore at high inclinations the same will be approximately true for $\varsigma$.

For very high inclinations it no longer makes sense to describe the Shapiro delay as the sum of a small number of harmonics (eq. 19). In this case it is better to use the exact expression for this sum given by:

$$\Delta_{\rm S}^{(3+)} = -2h_3 \left[ \frac{\ln(1+\varsigma^2 - 2\varsigma \sin \Phi)}{\varsigma^3} + \frac{2\sin \Phi}{\varsigma^2} - \frac{\cos 2\Phi}{\varsigma} \right]. \quad (28)$$

Adding the first and second harmonics we obtain:

$$\Delta_{\rm S} = -\frac{2h_3}{\varsigma^3} \ln(1+\varsigma^2 - 2\varsigma \sin \Phi), \quad (29)$$

which is the equivalent of eq. (9) re-written in the $h_3, \varsigma$ parameterisation.

## 4 TESTING THE NEW TIMING MODELS

To test our timing model, we used the TEMPO[1] software package. We modified the ELL1 timing model by including 3 new timing models based on the $(h_3, h_4)$ and $(h_3, \varsigma)$ (henceforth the "orthometric") parameterisations. Their properties are listed in Table 1 and their relative merits are discussed in § 4.1.

For high inclinations we preferably use the $h_3, \varsigma$ parameters. For low inclinations $\varsigma$ becomes ill-defined, making the $1/\varsigma^3$ factor used in the exact descriptions of the Shapiro delay potentially very large. This can cause the numerical routine that finds the best fit to diverge. In this case we use the $h_3, h_4$ parameterisation: eq. (19) is used to quantify $\Delta_{\rm S}^{(3+)}$ with a small number of harmonics and $\varsigma$ expressed as $h_4/h_3$.

In order to test these models and compare their strengths we created several sets of fake barycentric TOAs for a binary system with $e = 0$, $P_b = 12.3$ days, $m_p = 1.5 M_\odot$ and $m_c = 0.2 M_\odot$ (similar to PSR B1855+09 (Kaspi et al. 1994)) using the "exact" DD model (Damour & Deruelle 1986). This system is then "observed" at different orbital inclinations for about ten years (obtaining ten TOAs every fortnight) with a timing accuracy of 1 $\mu$s per TOA. We then use TEMPO to fit a timing model to these TOAs; using the timing models listed in Table 1.

---

[1] http://www.atnf.csiro.au/research/pulsar/tempo/

| Model | $\Delta_{\text{RS}} =$ | Harmonics | PK Parameters | Application |
|---|---|---|---|---|
| a | $\Delta_{\text{R}}$ (eq. 6) + $\Delta_{\text{S}}$ (eq. 9) | All (Exact) | $(r, s)$ | high inclinations |
| b | $\Delta_{\text{R}}$ (eq. 6) + $\Delta_{\text{S}}$ (eq. 29) | All (Exact) | $(h_3, \varsigma)^*$ or $(h_3, h_4)$ | high inclinations |
| c | $\Delta_{\text{RS}}^{(2)}$ (eq. 15) + $\Delta_{\text{S}}^{(3+)}$ (eq. 19) | 3 to $N$ (Approximate) | $(h_3, \varsigma)$ or $(h_3, h_4)^*$ | low inclinations |
| d | $\Delta_{\text{RS}}^{(2)}$ (eq. 15) + $\Delta_{\text{S}}^{(3+)}$ (eq. 28) | 3 to $\infty$ (Exact) | $(h_3, \varsigma)^*$ or $(h_3, h_4)$ | high inclinations |

**Table 1.** Propagational delays in four separate orbital models. Model "a" is ELL1, models "b", "c" and "d" are essentially the same except for the orthometric parameterisation of the propagational delays. These generally have a very low correlation between the parameters used to describe the Shapiro delay. The asterisk denotes preferred parameter pair. "Harmonics" specifies the harmonics used to describe the Shapiro delay, the description is "approximate" if harmonics above $N$ (a number specified by the user) are ignored. Models "c" and "d" have the advantage that the resulting parameters are not correlated to the Keplerian parameters $x$, $\kappa$ and $\eta$. Model "b" preserves the low correlation between $h_3$, $h_4$ (or $h_3$, $\varsigma$) but provides "true" values for $x$, $\kappa$ and $\eta$, at the expense of introducing a correlation with them. This model has the virtue that it can be readily extended to eccentric orbits.

### 4.1 Model comparison

We first compare the performance of the 3 exact high-inclination models for the $i = 75°$ data set. The results of this comparison are presented in Table 2. Whenever all harmonics of the Shapiro delay are used (models "a" and "b") the values of the Keplerian parameters are consistent with the values used in the simulation. As expected, there are strong correlations between these and the PK parameters.

If we use harmonics 3 and above to describe the Shapiro delay (model "d") then there are no correlations between the K and PK parameters, as expected. However, the values of the Keplerian parameters are modified by the 1st and 2nd harmonics of the Shapiro delay as predicted by eqs. (16) and (17). These modified values are also measured with better precision: the uncertainty of $x_p^{\text{obs}}$ is ten times smaller than that of than $x_p$ and $\eta^{\text{obs}}$'s is 3 times smaller than $\eta$'s. This happens because these "observed" parameters measure harmonic amplitudes, which can be determined directly from the TOAs; not computed quantities like $x_p$ and $\eta$ that are affected by uncertainties in the measurement of the Shapiro delay.

The most important lesson to retain from this comparison is the following: *the values of the astrophysically important quantities $(h_3, \varsigma)$ and their (small) cross-correlation are almost independent of whether we use all harmonics (model "b") or only the "measurable" harmonics (model "d") to describe the Shapiro delay*. More specifically, the improvement in the description of the Shapiro delay does not depend on the lack of correlations with the Keplerian parameters achieved in model "d". This means that the orthometric parameterisation is *intrinsically superior to the $r, s$ parameterisation*. This has important consequences, as we will see in § 6.1. In what follows we use only $\Delta_{\text{S}}^{(3+)}$ expressions to describe the Shapiro delay.

### 4.2 Comparing different inclinations.

We now compare the two parameterisations for different orbital inclinations. The PK parameters, their uncertainties and correlations are listed in Table 3.

Within their 1-$\sigma$ uncertainties, about 71% of the post-fit PK parameters ($r$, $s$, $h_3$ and $h_4$, $\varsigma$) match the values used to produce the TOAs. This is close to the 68% match rate that would generally be expected if the noise added is Gaussian. However, for $r$ and $s$ the uncertainties are very large for the lowest inclinations while the absolute uncertainties of $h_3$ and $h_4$ are almost constant: the absolute precision in the measurement of the amplitude of a harmonic is independent of the amplitude itself, it depends solely on the number and quality of the TOAs. Furthermore, $r$ and $s$ are very strongly correlated. In contrast, $h_3$ and $h_4$ are uncorrelated for the lower inclinations, as expected from the fact that they describe two orthogonal functions of the time delays.

At $i = 30°$ neither $r$ nor $s$ converge in a joint fit. In contrast, both $h_3$ and $h_4$ converge, but they are 2-$\sigma$ consistent with zero (see Table 3). The small $h_3$ in particular indicates that there is no conclusive detection of the Shapiro delay. Because of this $\varsigma \equiv h_4/h_3$ is not well defined; confirming that it is not a good parameter to use for the low orbital inclinations. Since $\varsigma$ is directly related to $s$ this explains why $s$ is not well defined either; this is the reason why the joint $r - s$ fit does not converge for this inclination.

The precision of $\varsigma$ improves faster with inclination than one would expect from the improved relative precisions of $h_3$ and $h_4$. As discussed in § 3.6, this happens because higher harmonics are contributing to the measurement of $\varsigma$, thus causing the strong positive correlation between $h_4$ and $h_3$ at high inclinations. It is also for this reason that the absolute uncertainties of $h_3$ and $h_4$ decrease, since they are mutually constrained by the precise measurement of their ratio. This causes an equalization of their *relative* uncertainties.

For very high inclinations we see a decrease in the correlation of $r$ and $s$. If we use $z_s = -\ln(1-s)$ instead of $s$ to parameterise the Shapiro delay (Kramer et al. 2006a) there is a further reduction in the correlation with $r$, as observed for PSR J0737−3039 (Kramer et al. 2006b). This means that for such edge-on systems the use of the $h_3, \varsigma$ parameterisation does not produce a significant improvement in the description of the constraints introduced by the Shapiro delay.

|  | Simulation Parameters | Fitted Timing Parameters | | |
|---|---|---|---|---|
|  |  | a | b | d |
| $x_p$ (lt-s) | 7.057488048268785 | 7.05748805(30) | 7.05748805(30) | - |
| $\eta$ | 0.0 | $-0.000000004(23)$ | $-0.000000004(23)$ | - |
| $x_p^{\rm obs}$ (lt-s) | - | - | - | 7.05749099(3) |
| $\eta^{\rm obs}$ | - | - | - | 0.000000329(8) |
| $r$ ($M_\odot$) | 0.2 | 0.187(26) | [0.1877] | [0.1935] |
| $s$ | 0.965925826289068 | 0.975(9) | [0.9746] | [0.9726] |
| $\varsigma$ | [0.7673] | - | 0.796(27) | 0.789(23) |
| $h_3$ ($\mu$s) | [0.4451] | - | 0.467(19) | 0.469(19) |
|  | Correlations | | | |
| $\rho(r,s)$ |  | $-0.96$ | - | - |
| $\rho(s,x_p)$ |  | 0.92 | - | - |
| $\rho(r,x_p)$ |  | $-0.99$ | - | - |
| $\rho(s,\eta)$ |  | 0.78 | - | - |
| $\rho(r,\eta)$ |  | $-0.90$ | - | - |
| $\rho(h_3,\varsigma)$ |  | - | $-0.50$ | $-0.51$ |
| $\rho(\varsigma,x_p)$ |  | - | 0.92 | $-0.03$ |
| $\rho(h_3,x_p)$ |  | - | $-0.79$ | 0.01 |
| $\rho(\varsigma,\eta)$ |  | - | 0.78 | $-0.02$ |
| $\rho(h_3,\eta)$ |  | - | $-0.85$ | 0.01 |

**Table 2.** Comparison of the exact high-inclination models "a", "b" and "d" for $i = 75°$. In this and following tables $\rho(x,y)$ indicates the correlation between parameters $x$ and $y$. Whenever all harmonics of the Shapiro delay are used (models "a" and "b") the values of the Keplerian (K) parameters are consistent with the values used in the simulation; but there are strong correlations between them and the post-Keplerian (PK) parameters. If we only use the 3$^{\rm rd}$ and higher harmonics then there are no correlations between the K and PK parameters, but the value of $x_p$ is modified by the 1$^{\rm st}$ harmonic of the Shapiro delay and the value of $\eta$ is modified by the 2$^{\rm nd}$ harmonic of the Shapiro delay (eqs. 16 and 17). Note the higher precision of the modified Keplerian parameters. -: was not fitted. In square brackets: PK parameters calculated from the PK parameters actually being fitted.

| Values used in Simulation | | | | | Values from "a" model | | | Values from new orbital models | | | | |
|---|---|---|---|---|---|---|---|---|---|---|---|---|
| $i$ (°) | $s$ | $\varsigma$ | $h_3$ ($\mu$s) | $h_4$ ($\mu$s) | $r/T_\odot$ ($M_\odot$) | $s$ | $\rho(r,s)$ | $h_3$ ($\mu$s) | $h_4$ ($\mu$s) | $\varsigma$ | $\rho(h_3,h_4)$ | $\rho(h_3,\varsigma)$ |
| 30 | 0.5 | 0.2679 | 0.0190 | 0.0051 | * | * | * | 0.032(21) | $-0.001(28)$ | * | $-0.04$ | * |
| 45 | 0.70710678 | 0.4142 | 0.0700 | 0.0290 | 0.8(5) | 0.25(40) | $-0.99$ | 0.088(21) | 0.008(27) | 0.08(29) | $-0.03$ | $-0.24$ |
| 60 | 0.86602540 | 0.5774 | 0.1896 | 0.1095 | 0.33(11) | 0.80(7) | $-0.99$ | 0.228(20) | 0.120(19) | 0.53(9) | 0.13 | $-0.37$ |
| 75 | 0.96592583 | 0.7673 | 0.4451 | 0.3415 | 0.187(26) | 0.975(9) | $-0.96$ | 0.469(19) | 0.370(13) | 0.789(23) | 0.71 | $-0.51$ |
| 85 | 0.99619470 | 0.9163 | 0.7579 | 0.6945 | 0.200(8) | 0.9958(7) | $-0.89$ | 0.749(17) | 0.683(13) | 0.912(7) | 0.94 | $-0.55$ |
| 89 | 0.99984770 | 0.9827 | 0.9348 | 0.9187 | 0.198(4) | 0.99984(4) | $-0.74$ | 0.924(14) | 0.907(12) | 0.9821(22) | 0.99 | $-0.51$ |

**Table 3.** Parameters used in our TOA simulations and results of fits made using the "a" orbital models and the new orbital models. The "c" model is used for $i = 30°$ and the "d" model is used for all other inclinations. In the simulations we used $m_c = r/T_\odot = 0.2 M_\odot$. The asterisks indicate instances where there was no numerical convergence. Note the small change in the precision of the measurement of $h_3$ and $h_4$ with orbital inclination.

The 1-$\sigma$ bands corresponding to the PK parameters in Table 3 are displayed graphically in Figs. 2-5. For $r$ and $s$ the 1-$\sigma$ bands are displayed in orange; for the orthometric parameters they are displayed in blue. In the $(\cos i, m_c)$ space the latter are calculated using eqs. (20), (21) or (22). They are then translated into 1-$\sigma$ bands in the $(m_p, m_c)$ space using eq. (5).

### 4.3 Bayesian analysis

For these sets of TOAs, we made a Bayesian $\chi^2$ analysis in the $(\cos i, m_c)$ space using the conventional $r, s$ parameterisation, as described in (Splaver et al. 2002). The choice of $\cos i$ instead of $\sin i$ comes from the fact that a group of systems with random orbital orientations will have a uniform distribution of $\cos i$. For each $\cos i, m_c$ combination we calculate the corresponding values of $r, s$. These parameters are then kept fixed, the spin and Keplerian orbital parameters are fitted and the post-fit $\chi^2$ is recorded. The 2-D probability distribution function (pdf) has a density given by

$$p(\cos i, m_c) \propto \exp\left(\frac{\chi^2_{min} - \chi^2(r,s)}{2}\right), \quad (30)$$

where $\chi^2_{min}$ is the global minimum of $\chi^2(r,s)$. For the low inclinations, where the resulting 2-D pdf is very spread out, we use $0 < m_c < 1.2\ M_\odot$; for the higher inclinations we focus the mapping in a smaller area where the pdf is non-zero. The 2-D pdf is then translated into a 2-D pdf in the $(m_p, m_c)$ space. For the lower inclinations we restrict this to $0 < m_p < 10 M_\odot$.

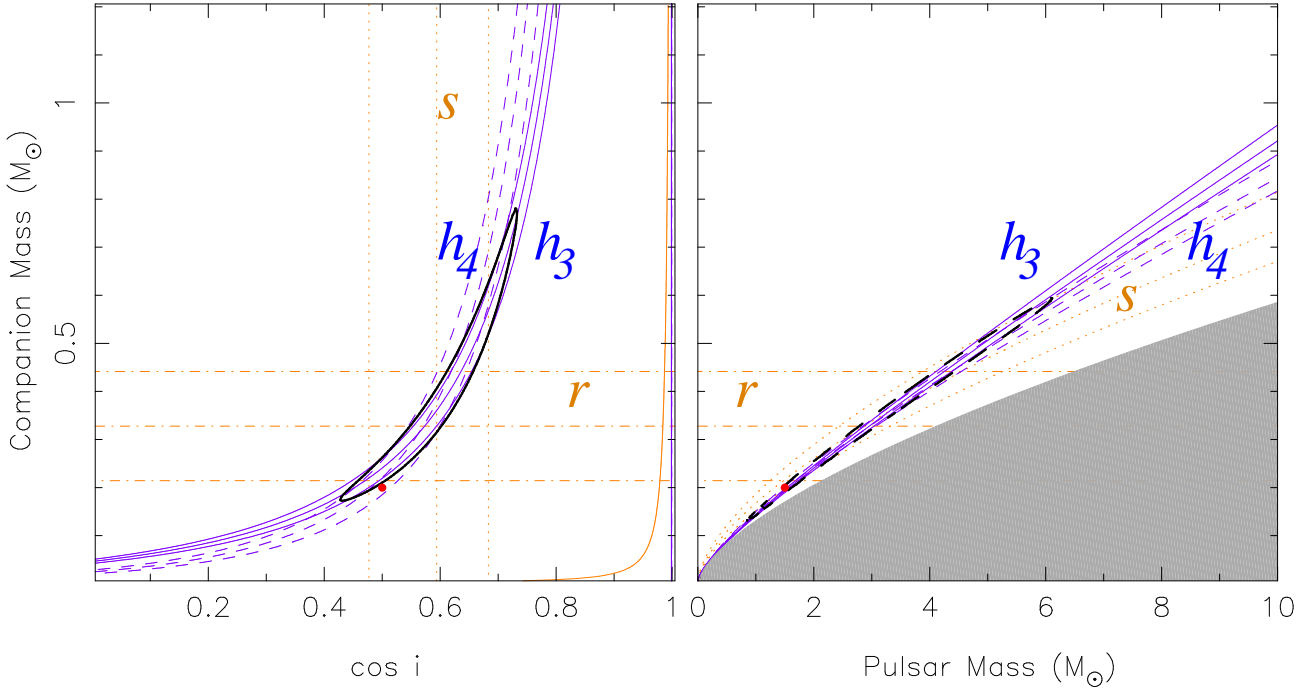

**Figure 2.** Shapiro delay constraints on the location of the PSR B1855+09 analogue with $i = 60°$ (high-inclination regime). For this and all following plots the red dots indicate the values for $m_p$, $m_c$ and $\cos i$ used to generate the simulated data. The orange curves indicate the 1-$\sigma$ bands corresponding to the values of $s$ (dotted) and $r$ (dot-dashed) listed in Table 3. The purple curves indicate 1-$\sigma$ bands of $h_3$ (solid) and $h_4$ (dashed). *Left*: $(\cos i, m_c)$ plot. In this and some of the following figures, the solid orange curve indicates $m_p = 0$. The black solid curve is a contour level of the 2-D probability density function that encloses 68.3% of the total probability. *Right*: $m_p - m_c$ plot. The black dashed curve is a contour level of the 2-D probability density function that encloses 68.3% of the total probability in this plane and within the range indicated in the figure. It is not a translation of the contour curve in the left plot. In this and following figures, the grey area is excluded by $\sin i \leqslant 1$. For this orbital inclination $h_3$ and $h_4$ are the best parameters to describe the Shapiro delay, since they are almost uncorrelated (Table 3). Although $r$ and $s$ are well-defined (they converge in a joint fit), the area of the intersection of their 1-$\sigma$ bands is much larger area than that enclosed by the 68% contours. Furthermore, the 1-$\sigma$ uncertainty for $r$ under-estimates the range of the pdf in $m_c$. In contrast, the intersection of the 1-$\sigma$ bands of $h_3$ and $h_4$ describes the allowed region very well. The shallow angle at which the 1-$\sigma$ bands of $h_3$ and $h_4$ meet is the reason why it is difficult to make precise mass estimates using Shapiro delay measurements.

The contours containing 68% of all probability in the $(\cos i, m_c)$ and $(m_p, m_c)$ spaces are superposed on the 1-$\sigma$ bands of the PK parameters in Figs. 2, 3, 4 and 5.

### 4.4 Discussion

One immediately noticeable feature of the 2-D pdfs displayed in Figs. 2, 3, 4 and 5 is their large extent to the high companion (and pulsar) masses, particularly for the low orbital inclinations. This is merely a consequence of the large uncertainty in the measurement of the companion mass for the low orbital inclinations, coupled with the truncation of the pdfs at $m_c = 0$, i.e., as the orbital inclination decreases and the uncertainty increases the pdf can only extend in one direction. Despite this, we can see that in 3 out of 4 cases (i.e. 75%) the parameters used to generate the simulated TOAs are within the contours that include 68% of the probability, which is as close to the expectation as possible given the small number statistics.

An inspection of these figures also shows that the orthometric parameters provide a much improved description of the $m_c$ and $i$ constraints derived from the Shapiro delay. Fig. 2 (high-inclination case) is particularly instructive: the 68% contour closely matches the intersection of the $h_3$ and $h_4$ bands, although it "prefers" the $h_3$ band since this is measured with better relative precision (see Table 3). $r$ and $s$ provide a very poor description of this 68% contour. This figure also provides a visualization of why is it so difficult to measure masses using the Shapiro delay (see § 3.5): the $h_3$ and $h_4$ curves cut each other at a very shallow angle, so small uncertainties in either $h_3$ or $h_4$ produce large differences in the $r$ where these curves intersect.

Fig. 3 provides a graphical description of the medium-inclination case. The 1-$\sigma$ band of $h_3$ (eq. 23) describes the location of the system very well; this also shows why $r$ and $s$ are so highly correlated for these lower orbital inclinations. Since $h_4$ is consistent with zero $\varsigma \equiv h_4/h_3$ must be small; this excludes high orbital inclinations (see § 3.4) but all the lower inclinations (up to arbitrarily large values of $m_c$) are allowed.

Fig. 4 provides a graphical description of the low-inclination case (§ 3.3), where the Shapiro delay is not de-tected. The values of $h_3$ and $h_4$ are consistent with zero within their 2-$\sigma$ uncertainties, therefore they overlap across all inclinations: it is for this reason that $s$ and $r$ no longer converge in a numerical fit. Despite that, it is still true that the system can only be in some parts of the $(\cos i, m_c)$ and $(m_p, m_c)$ spaces, as described by the 68% contours. The 1-$\sigma$ bands of $h_3$ and $h_4$ still provide a perfect description of these contours, i.e., useful constraints on the location of the system in the $(\cos i, m_c)$ space.

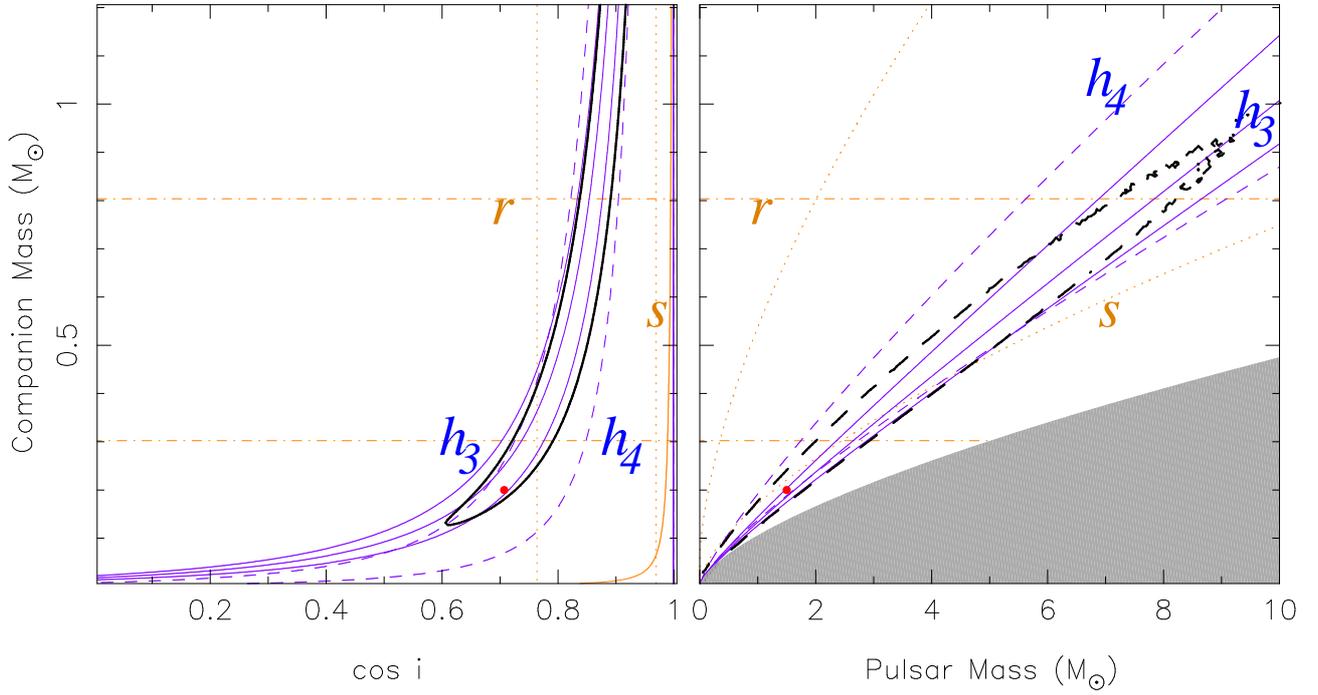

**Figure 3.** Probability distribution functions for $i = 45°$ (medium-inclination regime) displayed as in Fig. 2. For this inclination we can only measure the amplitude of the $3^{\rm rd}$ harmonic; $h_4$ is consistent with 0 (Table 3 - the 1-$\sigma$ lower limit of $h_4$ is not visible because it is negative). This excludes the high inclinations, but not the lower inclinations; the 2-D pdf now extends to arbitrarily large values of $m_c$ and $m_p$.

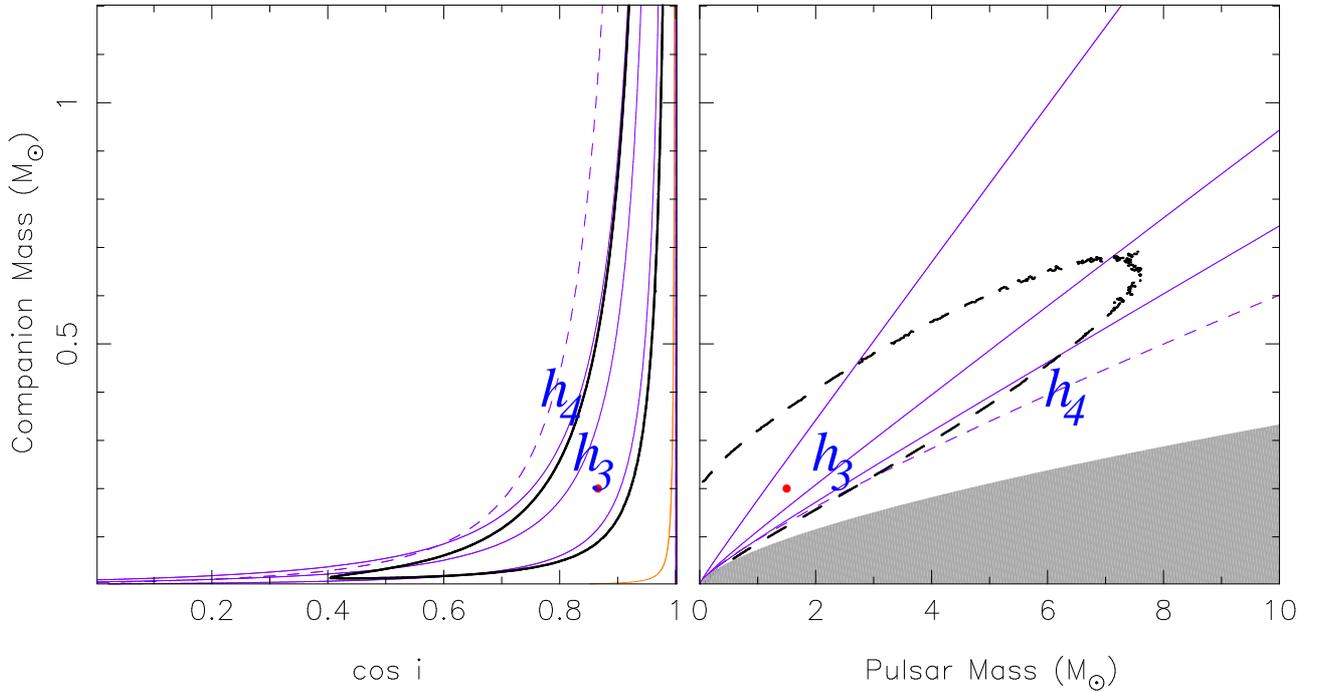

**Figure 4.** Probability distribution functions for $i = 30°$ (low-inclination regime), displayed as in Fig. 2. The Shapiro delay is not separable from the Rømer delay, i.e., $h_3$ and $h_4$ are now 2-$\sigma$ consistent with 0 (Table 3). Despite that, their 1-$\sigma$ bands still provide a perfect description of the 2-D pdf of the system, and represent useful constraints of its location in the $(\cos i, m_c)$ space. Note that for the small companion masses the system can now be located in a wide range or orbital inclinations; this is the reason why $s$ and $r$ no longer converge in a joint fit.

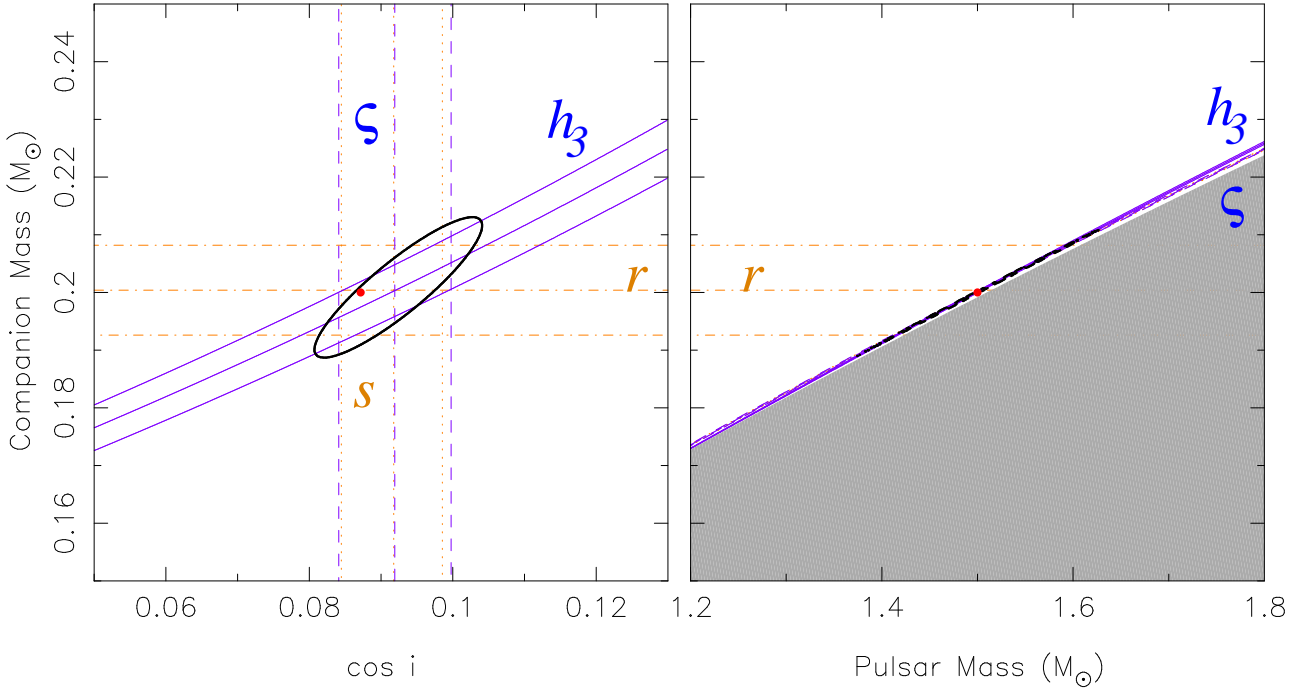

**Figure 5.** Probability distribution functions for $i = 85°$ (very high inclination regime), displayed as in Fig. 2 but for smaller inclination and mass ranges. For this inclination $h_3$ and $h_4$ are highly correlated, so $h_3$ and $\varsigma$ are the best parameters to describe the Shapiro delay.

Fig. 5 provides a description of the very high inclination case. For these high inclinations $\varsigma$ is the best parameter to describe the Shapiro delay (§ 3.6). In fig. 5 the 1-$\sigma$ band of $\varsigma$ covers almost the same region as the 1-$\sigma$ band of $s$, i.e., these two quantities are essentially interchangeable. This is to be expected because $\varsigma$ dependents on $s$ only (eq. 12). Despite the high precision in the measurement of $h_3$ and $\varsigma$, their respective curves still intersect at a very shallow angle in the $(m_c, m_p)$ plot; this explains the low precision of the measurement of $m_p$ even in cases where $h_3$ and $\varsigma$ (or $s$) are very precisely known.

Perhaps the most important lesson to be learned from these plots is the following: the total area of the 1-$\sigma$ bands of $h_3$ is not only smaller than those of $r$, but *smaller than those of s*. This applies particularly for the lower orbital inclinations (see Fig. 2). This has a very important consequence, discussed in § 6.1.

## 5 PROBABILITY MAPS IN ORTHOMETRIC SPACE.

We now concentrate on Fig. 3. One of the interesting features of this figure is that the region of high probability density extends to arbitrarily large values of $m_c$. For this particular orbital inclination ($i = 45°$) $h_4$ is consistent with zero but not $h_3$. This excludes high inclinations (see eq. 25) but allows for arbitrarily low inclinations, where the companion mass can be arbitrarily large (eq. 26). The 1-$\sigma$ bands of $h_3$ and $h_4$ (given by eqs. 23 and 24) thus have an infinitely larger overlap in the low inclination - high $m_c$ region than in the high-inclination - low $m_c$ region. Therefore, if we are making a $\chi^2$ map in the $(\cos i, m_c)$ space we end up with an arbitrarily large relative weight for the low inclinations and high masses, which will grow as we increase the range of $m_c$ being mapped.

We now suggest a way of deriving a distribution that, like the results of the $(r, s)$ fit, is independent of any specific upper limit of $m_c$. We begin by noting that whenever we make a $\chi^2$ map of the $(\cos i, m_c)$ space we implicitly assume an *a priori* constant probability density for $m_c$ and $\cos i$. It must be stressed that there is nothing unique about this choice. First, we could as well attribute an *a priori* constant probability to $m_p$. Because the translation between $m_c$ and $m_p$ is not linear (it is done with eq. 5) sampling the $(\cos i, m_p)$ space would produce a pdf that is different than that obtained by sampling the $(\cos i, m_c)$ space; our calculations show that it would be systematically offset to higher values of $m_p$ and $m_c$. Second, a uniform $\cos i$ (i.e., an assumption that the orbital orientations are *a priori* ran-domly aligned) is definitely a good starting assumption when no other information is available. However, in this case the timing provides further information regarding the orbital orientation and the mass of the companion.

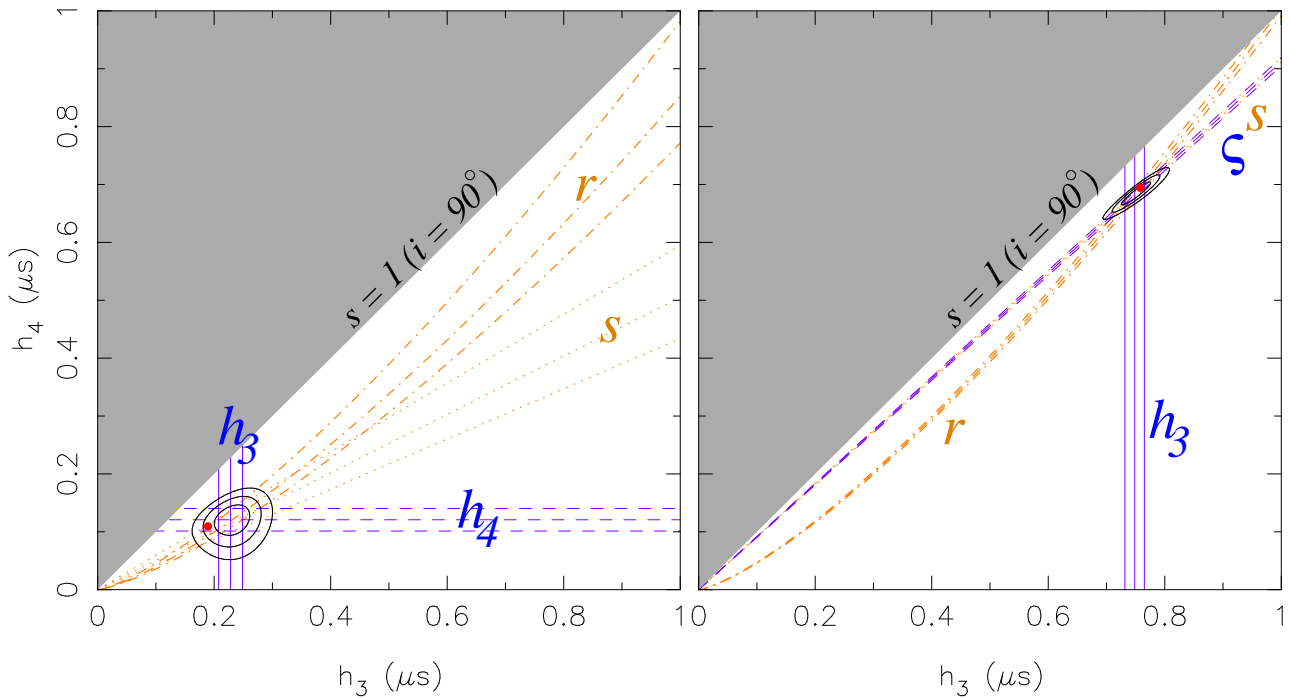

**Figure 6.** Shapiro delay constraints on the location of the PSR B1855+09 analogue in the orthogonal $h_3$-$h_4$ space. The red dots indicate the values for $h_3$ and $h_4$ that correspond to the $m_p$, $m_c$ and $\cos i$ used to generate the simulated data (see Table 3). The orange curves indicate 1-$\sigma$ bands of $s$ (dotted curve) and $r$ (dot-dashed). The smaller the value of $m_c$ the higher is its $r$ curve in this diagram, and the sooner it meets the $\sin i = 1$ line (where $\varsigma = 1$ and consequently $h_3 = h_4 = r$). The purple lines indicate the 1-$\sigma$ bands of $h_3$ (solid) and $h_4$ or $\varsigma$ (dashed) corresponding to the values in Table 3. The solid curves enclose 68.3%, 95.4% and 99.7% of the total probability. The grey area is excluded by the condition $\sin i \leqslant 1$. For $i = 60°$ (*Left*) $h_3$ and $h_4$ are *almost* uncorrelated (Table 3). The region near the $\sin i = 1$ line is less probable because it requires the presence of higher harmonics, which are not observed. This causes a small distortion in the 2-D pdf and its contours and introduces the small observed correlation. For $i = 85°$ (*Right*) $h_3$ and $h_4$ are significantly more correlated because the higher harmonics improve the precision of $\varsigma$.

We decide to map the $(h_3, h_4)$ space instead. For low inclinations these two parameters encode in an optimal way the timing information. Their orthogonality in this space eliminates the issue of the asymmetric overlap of their 1-$\sigma$ bands. Furthermore, we are making no direct assumptions about the *a priori* probability distributions of the physical parameters, we assume instead a constant *a priori* probability for the amplitude of the harmonics $h_3$ and $h_4$, which can be measured directly from the timing residuals.

We limit the mapping to the region where $h_4 \leqslant h_3$ (i.e., where $\sin i \leqslant 1$). We also restrict it to $m_c \geqslant 0$ which from eq. (26) implies $h_4 \geqslant 0$. We then use an exact timing model ("d") to fit for the spin and orbital parameters, but keeping $h_3$ and $h_4$ fixed. We then record the resulting $\chi^2$ and calculate probability maps as described in § 4.3.

The results of this mapping are presented in Fig. 6 for two inclinations ($i = 60°$ and $i = 85°$), using the same TOA datasets that were used to produce Figs. 2 and 5. The maps depict graphically some of the theoretical results described in § 3. For $i = 60°$ $h_3$ and $h_4$ are very weakly correlated, as one would expect from the fact that they represent the am-plitudes of two orthogonal functions. For $i = 85°$ the higher harmonics start improving the precision of the measurement of $\varsigma \equiv h_4/h_3$, this introduces a strong correlation between $h_3$ and $h_4$.

The "best fit" value of $r$ produced by the "d" orbital model ($r_b$) corresponds to the point where the lines for the nominal values of $h_3$ and $h_4$ meet: $r_b = h_3^4/h_4^3$. In these maps of the $(h_3, h_4)$ space there are very similar amounts of probability above and below $r_b$, despite the fact that the values of $r$ grow rapidly as we approach the $h_4 = 0$ line (at which point we have $r = +\infty$ and $\varsigma = s = 0$). The same will therefore be true when we re-project this 2-D pdf into the $(\cos i, m_c)$ and $(m_p, m_c)$ spaces, i.e., there will still be a similar amount of probability below and above $r_b$ despite the fact that there is infinitely more $r$ space above $r_b$.

In Fig. 7 we project the 2-D $(h_3, h_4)$ pdf for $i = 60°$ into the $(\cos i, m_c)$ and $(m_p, m_c)$ spaces (thick contours) and compare it with the 2-D pdf obtained by sampling the $(\cos i, m_c)$ space directly (displayed in Fig. 2 and again here with thin contours). Although in both cases the 68% contours can be reasonably approximated by the intersection of the 1-$\sigma$ bands of $h_3$ and $h_4$, they are different amongst themselves: the thin contours are skewed towards higher masses relative to the thick contours. Looking at the marginal 1-D pdf for $m_c$, we can also see that the 1-D pdf derived from the $\chi^2$ map of the $(h_3, h_4)$ space produces median and lower 1-$\sigma$ limits that are (at least in this case) in better agreement with the results of the $r, s$ fit (in orange) than those derived from the $\chi^2$ map of the $(\cos i, m_c)$ space.

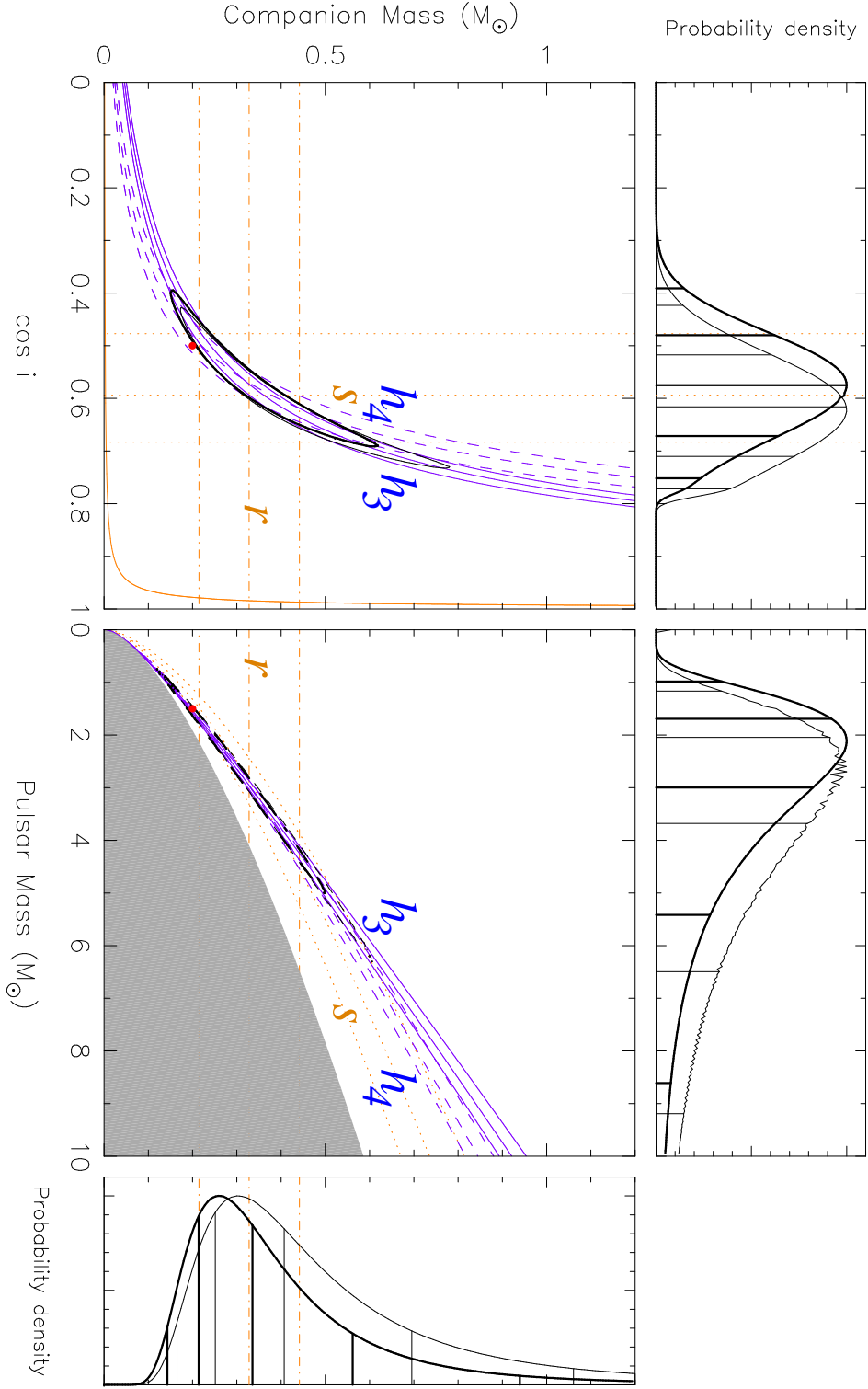

**Figure 7.** Same as Fig. 2, but now for two 2-D pdfs and with marginal 1-D probability distribution functions displayed in the marginal plots. The distance between the 0 and the vertical lines in these plots (horizontal in the $m_c$ plot) includes 2.28%, 15.86%, 50% (median), 84.14% and 97.72% of the total probability; these correspond to 1 and 2-$\sigma$ intervals around the median. The 1-D pdfs are not Gaussian, so the median and the peak in probability do not occur at the same locations. The thin contours correspond to the same pdf displayed in Fig. 2, obtained by mapping the probability in $r - s$ space. The thick contours are a projection of the 2-D pdf displayed on the left panel of Fig. 6 ($i = 60°$) in the $m_c$-$\cos i$ and $m_c$-$m_p$ spaces. The 1-D pdf obtained from projecting the 2-D pdf along the $m_c$ axis has a median and 1-$\sigma$ lower limits that are very similar to the values provided by the $r, s$ fit. We can also see that the former map has a systematic shift towards higher masses. This would be even more apparent were we to extend the mapping to higher values of $m_c$ and $m_p$.

Although it is problematic to make a $\chi^2$ map of the $(\cos i, m_c)$ space for low orbital inclinations (in some cases the probabilities depend sensitively on the upper $m_c$ limit of the map, see e.g. Fig. 3) this is not a problem for high orbital inclinations. In the latter case the overlap of the 1-$\sigma$ bands is nearly symmetrical relative to the "best" solution given by the intersection of $h_3$ and $\varsigma$ (see fig. 5), so there is no systematic offset between the two types of maps. Indeed, projecting the 2-D $(h_3, h_4)$ pdf for $i = 85°$ into the $(\cos i, m_c)$ and $(m_p, m_c)$ spaces we see no noticeable difference in the resulting 2-D and marginal 1-D pdfs.

# 6 IMPROVED TEST OF GENERAL RELATIVITY

As discussed in § 2, if we can determine more than two PK parameters for a binary pulsar the system of equations used to solve $m_c$, $m_p$ and $i$ becomes over-determined and we can test general relativity. A good example of this is PSR B1534+12, a double neutron star system with a relatively compact ($P_b = 10.1$ hr) and eccentric ($e = 0.27$) orbit. This allowed for the first time the measurement of a total of 5 PK parameters (Stairs et al. 1998; Stairs et al. 2002). In this system, the two PK parameters measured with better precision ($\dot{\omega}$ and $\gamma$) can be used to determine $m_c$ and $m_p$, assuming GR to be the correct theory of gravity. Once this is done, the remaining 3 PK parameters represent potential tests of general relativity. The observed orbital period decay ($\dot{P}_b$), which is mostly due to loss of orbital energy due to the emission of gravitational waves, has not provided an interesting test for this particular system because its distance is not well known. This precludes an accurate correction of the kinematic contributions to the observed $\dot{P}_b$, particularly the contribution due to the Shklovskii effect (Stairs et al. 2002). However, the relatively high orbital inclination of the system ($\sim 77.2°$), large companion mass ($1.3452 \pm 0.0010$ $M_\odot$ (Stairs et al. 2002)) and relatively good timing precision (4 to 6 $\mu$s) have allowed a highly significant detection of the Shapiro delay. The two PK parameters that describe it provide two independent tests of GR. In this and other similar cases an improved parameterisation of the Shapiro delay should lead to improved parametric tests of GR.

## 6.1 Eccentric orbits

The "precise" PK parameters ($\dot{\omega}$ and $\gamma$) can only be measured if the orbit is eccentric. Therefore, if we are to use the orthometric parameterisation to improve the precision of GR tests we need to find out first whether it is also a better description of the Shapiro delay for eccentric orbits.

The basic assumption made at the start of § 3 was that the first and second harmonics of the Shapiro delay can be completely absorbed in the Rømer delay, and that the higher harmonics are due solely to the Shapiro delay. For eccentric orbits this is no longer the case, particularly if there is a significant change in the longitude of periastron over the time span of observations. Moreover, the Fourier expansion of the Shapiro has additional terms that are proportional to $r$ and independent of the inclination of the orbit: in an eccentric binary there is a phase-varying Shapiro delay even if it is seen face-on ($i = 0°$). For this reason, and also because of their inherent simplicity, we will from now on only consider exact descriptions of the Shapiro delay, like eq. (9) for the near circular case or eq. (2) for the general elliptical case. Dropping constant terms the latter equation can be re-written as a function of $h_3$ and $\varsigma$:

$$\Delta_S = \frac{2h_3}{\varsigma^3} \left[\ln(1 + e\cos\varphi) - \ln(1 + \varsigma^2 - 2\varsigma\sin(\omega + \varphi))\right]. \quad (31)$$

Note that this equation has the same limitation as eqs. (28) and (29): it can only be used when $\varsigma$ is a well-defined quantity, i.e., for high orbital inclinations. Furthermore, in this case the $h_3$ no longer has the simple physical interpretation (amplitude of the third harmonic) that it had for low-eccentricity orbits.

In § 4.1 we found that, despite introducing correlations between $h_3$, $\varsigma$ and the Keplerian parameters, the use of the exact expression for the Shapiro delay with all harmonics (eq. 29, used in model "b") preserves the low correlation between $h_3$ and $\varsigma$ obtained using solely the higher harmonics (eq. 28, used in model "d"), which is much lower than $\rho(r, s)$ in the traditional parameterisation (model "a"). In what follows, we verify whether the latter proposition is also true in the elliptical case, i.e., whether despite the use of all harmonics the orthometric parameterisation still provides an improved description of the Shapiro delay.

## 6.2 Implementing and testing the eccentric timing model

The DD model now distributed in TEMPO uses eq. (2) to describe the Shapiro delay, with $r, s$ as parameters. We extended this orbital model with the option of using eq. (31) instead, with $h_3$ and $\varsigma$ as parameters. The latter is an eccentric analogue of model "b" in Table 1, since it uses an exact expression for the Shapiro delay with all the harmonics that is a function of $h_3$ and $\varsigma$.

To test this orbital timing model, we created a list of TOAs for a simulated pulsar with orbital parameters similar to those of PSR B1534+12. This "pulsar" was "observed" every two weeks for a period of 20 years. In each observation we obtain 4 TOAs each with an uncertainty of 5 $\mu$s. We fit the traditional DD model to these TOAs using the $r, s$ parameterisation for the Shapiro delay. The results of this fit are presented in Table 4. We then make the same fit using the orthometric parameterisation for the Shapiro delay; the results are also displayed in Table 4.

We also made a $\chi^2$ map of the $(\cos i, m_c)$ space: as mentioned above, for these high inclinations there is no systematic offset between the mass distributions produced in the $\cos i, m_c$ and $h_3, h_4$ spaces. For each point we keep only the corresponding values of $r$ and $s$ fixed and allow all other parameters (including the remaining PK parameters) to vary freely. We then record the resultant $\chi^2$ and calculate the 2-D pdf as discussed in § 4.3. In this manner we can see the real constraints on the location of the system in the $(\cos i, m_c)$ space derived from the Shapiro delay.

| Parameter | Simulation | $(r, s)$ | $(h_3, \varsigma)$ | Value / GR prediction |
|---|---|---|---|---|
| Timing Parameters | | | | |
| $m_p$ ($M_\odot$) | 1.3332 | | | |
| $m_c$ ($M_\odot$) | 1.3452 | | | |
| $i$ (°) | 77.2 | | | |
| $P_b$ (days) | 0.420737299153 | 0.420737299158(4) | 0.420737299158(4) | |
| $x_p$ (lt-s) | 3.729511453142 | 3.7295120(12) | 3.7295120(12) | |
| $e$ | 0.2736767 | 0.27367662(14) | 0.27367662(14) | |
| $\omega$ (°) | 274.76928 | 274.76925(4) | 274.76925(4) | |
| $\dot{\omega}$ (°yr$^{-1}$) | [1.7557682] | 1.755770(3) | 1.755770(3) | |
| $\gamma$ (s) | [0.0020696] | 0.0020680(8) | 0.0020680(8) | |
| $r$ ($M_\odot$) | [1.3452] | 1.35(11) | [1.35276] | $1.00 \pm 0.08$ |
| $s$ | [0.975149] | 0.974(4) | [0.97416] | $0.999 \pm 0.004$ |
| $h_3$ ($\mu$s) | [3.370681013] | - | 3.34(11) | $0.99 \pm 0.03$ |
| $\varsigma$ | [0.798289512] | - | 0.794(15) | $0.996 \pm 0.018$ |
| Correlations | | | | |
| $\rho(r, s)$ | | $-0.96$ | - | |
| $\rho(r, \dot{\omega})$ | | $-0.49$ | - | |
| $\rho(r, \gamma)$ | | $-0.12$ | - | |
| $\rho(r, x_p)$ | | $-0.95$ | - | |
| $\rho(r, e)$ | | $-0.78$ | - | |
| $\rho(s, \dot{\omega})$ | | $+0.43$ | - | |
| $\rho(s, \gamma)$ | | $+0.12$ | - | |
| $\rho(s, x_p)$ | | $+0.87$ | - | |
| $\rho(s, e)$ | | $-0.65$ | - | |
| $\rho(h_3, \varsigma)$ | | - | $-0.69$ | |
| $\rho(h_3, \dot{\omega})$ | | - | $-0.49$ | |
| $\rho(h_3, \gamma)$ | | - | $-0.09$ | |
| $\rho(h_3, x_p)$ | | - | $-0.88$ | |
| $\rho(h_3, e)$ | | - | $+0.84$ | |
| $\rho(\varsigma, \dot{\omega})$ | | - | $+0.43$ | |
| $\rho(\varsigma, \gamma)$ | | - | $+0.12$ | |
| $\rho(\varsigma, x_p)$ | | - | $+0.87$ | |
| $\rho(\varsigma, e)$ | | - | $-0.65$ | |

**Table 4.** Comparison of $r, s$ and $h_3, \varsigma$ parameterisations of the Shapiro delay for an eccentric orbit. In the first column, the values in square brackets are derived from $m_p$, $m_c$, $i$ and the Keplerian parameters. In the third column, $r$ and $s$ are derived from the values of $h_3$ and $\varsigma$

The 1-$\sigma$ bands corresponding to the PK parameters in Table 4 are displayed graphically in Fig. 8. On top of these we overlay the contour of the that 2-D pdf that includes 68.3% of its total probability.

### 6.3 Discussion

The results in Table 4 show that the computation of the Keplerian and post-Keplerian parameters (apart from those associated with the Shapiro delay) are not affected at all by the choice of the parameterisation of the Shapiro delay. This is expected from the exact equivalence of the delays predicted by the two parameterisations (eq. 2 and eq. 31).

Second, not much changes in terms of correlations with the remaining Keplerian and post-Keplerian parameters, except that, as in the circular case, $h_3$ and $\varsigma$ are less correlated with each other than $r$ and $s$. This shows that the orthometric parameterisation is also a superior description of the Shapiro delay for eccentric orbits.

Third, we can use the values of $m_p$, $m_c$ and $i$ used to simulate the TOAs to predict $s$, $\varsigma$, $r$ and $h_3$ using the equations in § 3; these predictions are presented in the first column of Table 4. For a real pulsar we would instead use the measurements of $\dot{\omega}$, $\gamma$ and $f$ to calculate $m_p$, $m_c$ and $i$ assuming that general relativity is correct and then predict $s$, $\varsigma$, $r$ and $h_3$. At the level of precision to which the Shapiro delay parameters are measured it does not make any difference which method we use make this prediction.

For the simulated set of PSR B1534+12 TOAs GR passes both the traditional $r, s$ tests and also the new $h_3, \varsigma$ tests. Note, however, that the $h_3$ test is more stringent than the previous best test ("$s$"): by this we mean that for any particular value of $\dot{\omega}$ or $\gamma$, the measurement of $h_3$ and its uncertainty implies a smaller range of $\cos i$, $m_c$ and $m_p$ than that allowed by the measurement of $s/\varsigma$ (see Fig. 8).

## 7 CONCLUSIONS

In this paper we discussed what can be learned from an expansion of the Shapiro delay in harmonics of the orbital period for a system with small orbital eccentricity, as is the case for the vast majority of the millisecond pulsars where this effect can be measured.

The amplitudes of these harmonics, measured directly from the time delays, provide a much improved parametric description of the regions of the $m_c$-$\cos i$ space where the binary can be located, even when no Shapiro delay can be measured. In particular, we show that the orthometric amplitude parameter $h_3 = r \left( \frac{s}{1+\sqrt{1-s^2}} \right)^3$ is always measured with significantly higher precision than one would generally estimate from the uncertainties of $r$ and $s$. For low inclinations, the amplitude of the fourth harmonic ($h_4$) is the best parameter to complete the description of the Shapiro delay because $h_3$ and $h_4$ are then uncorrelated. For high inclinations the amplitudes of the higher harmonics improve the precision of the measurement of the orthometric ratio parameter $\varsigma \equiv h_4/h_3$ beyond what is possible from the individual values of $h_3$ and $h_4$; therefore the latter become strongly correlated. If $|\rho(h_3, h_4)| > |\rho(h_3, \varsigma)| = 0.5$ the latter become the best parameters to describe the Shapiro delay. Because of these low correlations, the new orthometric parameters provide superior parametric descriptions of the mass and orbital inclination constraints determined from the Shapiro delay, which previously could only be described by probabilistic $\chi^2$ maps of the $(\cos i, m_c)$ space.

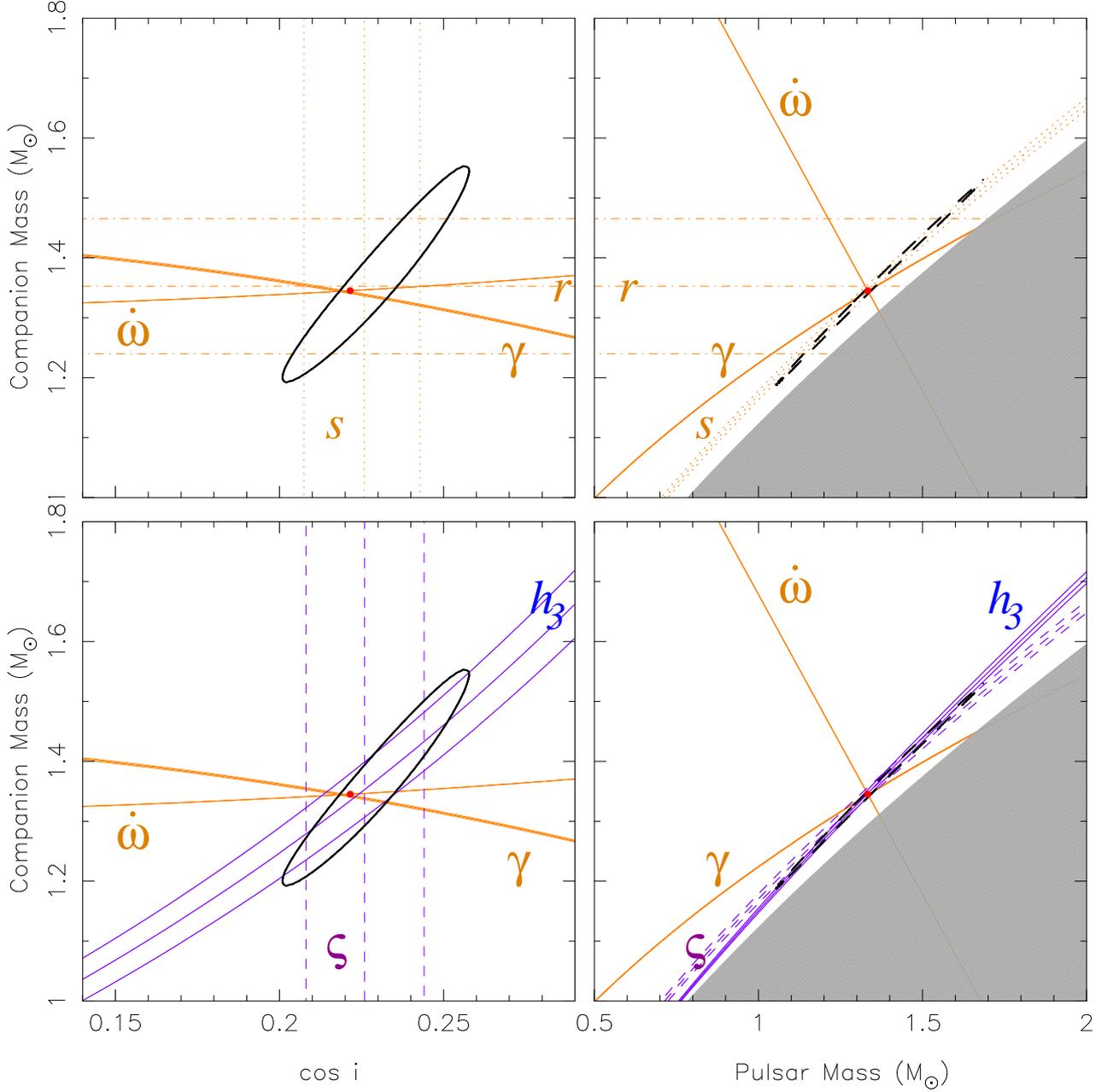

**Figure 8.** Constraints on the location of the PSR B1534+12 analogue derived from four PK parameters. The orange curves indicate the 1-$\sigma$ bands corresponding to the values of traditional PK parameters listed in Table 4. The purple curves indicate the 1-$\sigma$ bands of $h_3$ (solid) and $\varsigma$ (dashed), also listed in that Table. The black solid contours enclose 68.3% of the total probability in the 2-D pdf. *Left*: $(\cos i, m_c)$ plot; *Right*: $m_p - m_c$ plot, with the grey area excluded by $\sin i \leqslant 1$. *Top*: The current $r, s$ tests of general relativity. *Bottom*: New $h_3, h_4$ tests of general relativity. The mass function and the intersection of the 1-$\sigma$ bands of $\dot{\omega}$ and $\gamma$ produce estimates of $i$, $m_c$ and $m_p$ which are close to the value used in the simulation (red dots). As in the circular case, the 1-$\sigma$ bands of $h_3$ and $\varsigma$ provide a superior description of the 63.8% contours allowed by the Shapiro delay. The $s$ and $\varsigma$ tests are essentially equivalent: their 1-$\sigma$ bands are nearly identical, despite the apparent difference in relative precision (see Table 4). On the other hand, the $h_3$ test (which essentially replaces the $r$ test) is much more stringent. Remarkably, it is also more stringent than the $s/\varsigma$ test: for any particular value of $\dot{\omega}$ or $\gamma$ it allows smaller ranges of $m_c$, $m_p$ and $i$.

The $\chi^2$ maps of the $(h_3, h_4)$ space make no explicit assumptions about the *a priori* probability distributions of the physical parameters, we assume instead a constant *a priori* probability for $h_3$ and $h_4$, which can be measured directly from the timing residuals. Unlike in the case of a $\chi^2$ maps of the $(\cos i, m_c)$ space, they don't depend on the range of $m_c$ being mapped and guarantee that the total probability is nearly equally split between regions with $r$ above and below the "best" value $r_b$.

We also show that the improved description of the Shapiro delay can be extended to high-eccentricity binaries. This has the consequence that in eccentric systems where other PK parameters are known we can now make a significantly improved test of GR by measuring $h_3$ instead of $r$. Remarkably, the $h_3$ test is more stringent than the previous $s$ test.

It is important to note that no new information is provided by this re-parameterisation of the Shapiro delay beyond what was previously provided by a $\chi^2$ map. Its main advantage is being concise, since we only need two numeric parameters to describe the 2-D pdf. Both techniques extract information from the observed timing delays that was not provided by the previous $r, s$ parameterisation.

**ACKNOWLEDGEMENTS**

We thank Marten van Kerkwijk for his comments and suggestions regarding this work. Since completing it we have learned that he has independently understood the need for an improved parameterisation of the Shapiro delay and, in an unpublished work, derived some of the results discussed here, namely the Fourier expansion of the Shapiro delay and also an equivalent of eq. (28) albeit in a slightly different non-orthogonal parameterisation. We also thank the referee, Ingrid Stairs, for her careful and meticulous review and thoughtful suggestions, which have improved the quality of this work.